# Analysing the Social Spread of Behaviour: Integrating Complex Contagions into Network Based Diffusions

Josh A. Firth*[†,1,2], Gregory F. Albery[3,4], Kristina B. Beck[1,5], Ivan Jarić[6,7], Lewis G. Spurgin[8], Ben C. Sheldon[1] & Will Hoppitt*[†,9]

[1]Department of Zoology, Oxford University, Oxford, UK
[2]Merton College, Merton Street, Oxford, UK
[3]Department of Biology, Georgetown University, Washington, DC
[4]Institute of Evolutionary Biology, University of Edinburgh, Edinburgh, UK
[5]Department of Behavioural Ecology and Evolutionary Genetics, Max Planck Institute for Ornithology, Seewiesen, Germany.
[6]Biology Centre of the Czech Academy of Sciences, Institute of Hydrobiology, České Budějovice, Czech Republic,
[7]University of South Bohemia, Faculty of Science, Department of Ecosystem Biology, České Budějovice, Czech Republic
[8]School of Biological Sciences, University of East Anglia, Norwich, UK
[9]School of Biological Sciences, Royal Holloway, University of London, Egham, UK

*Correspondence: Joshua.firth@zoo.ox.ac.uk & whoppitt@mac.com  [†]Joint lead authors

**ABSTRACT**
The spread of socially-learnt behaviours occurs in many animal species, and understanding how behaviours spread can provide novel insights into the causes and consequences of sociality. Within wild populations, behaviour spread is often assumed to occur as a 'simple contagion'. Yet, emerging evidence suggests behaviours may frequently spread as 'complex contagions', and this holds significant ramifications for the modes and extent of transmission. We present a new framework enabling comprehensive examination of behavioural contagions by integrating social-learning strategies into network-based diffusion analyses. We show how our approach allows determination of the relationship between social bonds and behavioural transmission, identification of individual-level transmission rules, and examination of population-level social structure effects. We provide resources that allow general applications across diverse systems, and demonstrate how further study-specific developments can be made. Finally, we outline the new opportunities this framework facilitates, the conceptual contributions to understanding sociality, and its applications across fields.

**INTRODUCTION**

Social learning allows the transmission of information between individuals, and is widespread across the animal kingdom (*1, 2*). Evidence of social transmission has been provided across the range from highly social to largely solitary species (*3, 4*), and from cognitively complex species to those assumed to be less cognitively developed (*1, 5*). New insights are continually being generated into the proximate and evolutionary causes of social learning (*2, 6*), and its ecological consequences (*7*). Importantly, socially transmitted information can change an individual's behaviour, which can in turn be observed and copied by others, meaning the behaviour can spread through the population (*8*). Some of the first observations of this phenomenon include the apparent social transmission of opening milk bottle tops by British tits (*Paridae*) (*9*), and Japanese macaques spontaneously learning to wash sweet potatoes and subsequently spreading this behaviour (*10*). Yet, although social learning processes have therefore long been recognized, the methods used to understand the mechanism of behavioural transmission are still developing.

Fundamentally, individuals can only learn from those they hold some kind of social connection to. As such, social transmission is by definition dependent upon fine-scale social connections within a population. Recent advances in observing and quantifying the networks of social connections that occur within animal systems - i.e. 'animal social networks' (*11*) - have allowed the study of social transmission to move away from simply assessing demonstrator-observer paradigms within dyads, or learning curves across the population. Instead, contemporary techniques examine how socially-informed behaviours spread through the social networks of natural populations (*12, 13*).

Network-Based Diffusion Analysis (NBDA) is a widely-applied framework that allows researchers to assess how social networks determine the diffusion of socially transmitted behaviours. Specifically, it provides a way of quantifying the extent to which the timing or order of adoption of socially learnt behaviours by individuals is related to an observed social network (*13*), and -importantly- outputs a parameter ('*s*') that estimates the likelihood of uninformed individuals (i.e. those that haven't yet adopted the behaviour) adopting a behaviour given a single unit of social connection to 'informed' individuals (i.e. those that have already adopted the behaviour). For example, an *s* value of 2 would mean that an uninformed individual with single unit of social connection to an informed individual is twice as likely to learn a behaviour next compared to an uninformed individual with no connections to informed individuals. In this way, this general approach has been used to investigate how networks shape the social spread of information, possible transmission pathways, and the importance of social *versus* asocial learning in shaping the adoption of behaviours (*14, 15*). Largely due to its power and flexibility, NBDA has now been employed across species from captive insects and fish (*16, 17*) to wild birds and mammals (*18-20*). Furthermore, the framework has recently been expanded to control for individual-level differences in asocial and social learning (*13, 21*), to compare different types of networks (*14, 19*), and to include dynamically changing networks (*22*), under both frequentist and Bayesian settings (*23*).

Although NBDA provides a highly versatile platform for assessing and quantifying social diffusion through animal networks (*15*), its methodology makes some important assumptions that have to be addressed. Like most animal social network studies, a central aspect of the current NBDA framework commonly assumes that the probability that an 'uninformed' individual socially learns a new behaviour is directly dependent on the number and strength of connections (within the given network) that the individual holds to 'informed' others that are currently displaying the behaviour (*12-14*). This is equivalent to assuming that the relevant exposure to a behaviour determines an uninformed individual's opportunity to learn that behaviour themselves. As such, the diffusion of the behaviour is - in this way - treated as analogous to the spread of a contagious disease, whereby infection probability is generally determined by the amount of relevant contacts with infected individuals, which is often referred to as a 'simple contagion' (*24, 25*).

Essentially, the simple contagion process views the effect of each dyadic connection within the network as somewhat independent from the other dyadic connections, so that every connection an uninformed individual has to informed individuals holds a set probability of allowing social transmission, regardless of the other connections the uninformed individual has within that network (e.g. to other uninformed individuals or informed individuals). A particularly appealing property of potentially considering behavioural spread as a simple contagion on a network is that it would appear to allow for a universal concept of diffusion across different spreading processes. If this were true, it would mean that the established ideas developed by contagious disease epidemiology models could be directly applied to understanding social transmission of behaviours, and therefore it is perhaps unsurprising that this assumption (of the simple contagion of behaviours) was highly influential within sociology (i.e. the field it was first developed) (*26*) and also has been frequently drawn upon when considering animal social systems (*27*).

Despite the influence and appeal of considering the spread of behaviours as simple contagions, the utility of this approach as a universal model to understand the spread of information in real-world systems has been increasingly questioned in the field it was first established (sociology) (*26*), and thus subsequently in animal systems (Firth 2020 TREE). Most convincingly, recent theoretical developments and mounting empirical evidence from human systems has demonstrated that considering socially-transmitted behaviours as simple contagions is inaccurate, and that 'disease spread' may be a poor, or even somewhat opposite, analogy for behavioural contagion (*26, 28*). Specifically, sociological research across topics as diverse as health behaviours (*29, 30*), political beliefs (*31*), and the spread of innovations (*32*) has repeatedly shown weak support for considering each dyadic social connection to independently influence the adoption of a behaviour (such that even single connections may be sufficient to cause behavioural change), and has instead found substantial evidence for 'complex contagions' (*26, 33*).

Like simple contagions, complex contagions follow the basic principle that social transmission occurs between individuals that are socially connected to one another. But, unlike simple contagions, complex contagion processes involve some 'complexity', in that the likelihood of adopting the behaviour depends on more than just the raw amount of social ties to informed individuals. Instead, the adoption of the behaviour may require social reinforcement via multiple contacts, or be related to an uninformed individual's proportion of social connections to informed individuals (rather than the absolute amount of social connections to them) (*25, 34*).

Considering complex rather than simple contagions can radically alter predictions of behavioural spread (*26, 27*) (Figure 1). Firstly, at the individual level, under a simple contagion we would assume that the most socially central individuals will be likely to quickly adopt a behaviour (due to being likely to hold links to some informed individuals), and be highly important in spreading a behaviour (e.g. as often assumed across the animal behaviour literature (*35*)) (Figure 1A). But, under complex contagion, the most central individuals may be late adopters (e.g. due to many links to non-informed individuals) and also be less influential in spreading novel behaviour (e.g. due to late adoption, and because they are connected to other, highly-connected, individuals) (*27*) (Figure 1B). Conversely, at the group-level, highly clustered cliques should be relatively unimportant for simple contagions due to their low likelihood of connections to those expressing a new behaviour (until a group member adopts) and also then the high 'redundancy' of social ties – most of which are within their own group, limiting further spread (*24*). Yet, these tightly banded groups are often crucial for promoting complex contagions, as strong ties within clustered cliques become most important when spreading requires social reinforcement (*36, 37*). Finally, at the population-level, social network structures with lots of weak links between widely separated individuals and groups are likely to promote simple contagions by allowing spreading to occur across these 'bridges' (*38*), while complex contagions will occur faster in structures with higher clustering (and fewer long-bridges) that promotes social reinforcement and cascading within cliques (*26*).

Despite the acknowledged importance of complex contagions across sociology (*26*) and emerging evidence for their role in human behaviour (*28, 39*), the concept remains largely neglected within non-human animal behaviour (*27*), even though early lines of evidence indicate it is similarly important in these systems (*27, 36, 40*). In this paper, we develop a method of assessing diverse forms of complex contagions in generalisable ways, and integrate this concept into NBDA (the most prevalent approach for considering social transmissions in real-world social networks) to allow researchers to move beyond considering spread within a simple contagion framework, and instead to examine the evidence, the extent, and the relative importance, of complex contagions and social learning strategies in social networks.

Specifically, (1) we first conceptually extend NBDA to a generalised framework allowing the fitting of complex contagions. We then (2) specify three possible (and commonly considered) transmission rules, and in each case we demonstrate how to enter the data, fit the model, simulate relevant example/test data sets, and interpret the model output. Further, using these different transmission rules, we (3) directly test the performance of the new method based on how successful the complex NBDA is at estimating the true relationship between an individuals' social connections and their chance of adopting a behaviour (i.e. the user-inputted *s* value), and at recovering the transmission rule in operation (i.e. the simulated social learning strategy). We also provide the resources for researchers to test these methods' performance for their own transmission rules, specifically in the context of their own empirical data (i.e. of different sample sizes and network structures). Our simulations of social networks and various social contagion types demonstrate how this new method is superior for assessing the pathways of social transmission, and mode of contagion, under a range of intuitive scenarios. This framework provides a new resource to enable generalised assessment of behavioural spread, while also allowing flexibility (e.g. in terms of accounting for individual-level differences in asocial innovation and comparison of contagions across types of networks). Finally, we discuss how this method can be used for various applications, and how future extensions or developments may allow continued insight into the consequences of sociality.

**METHODS AND RESULTS**

The standard NBDA method and various extensions of this have been heavily applied in behavioural science, and described in various user-type guides and reviews, most recently Hasenjager et al. 2020(*15*). We developed NBDA methods for considering complex contagions, and explored the results in three stages. All of the resources are freely available at https://github.com/whoppitt/complexNBDA. First, we *(1)* generalise the NBDA method to allow assessment of complex contagions (Figure 1), and provide three intuitive specifications of different usable transmission rules as *(1i) Proportional rule model (1ii) Frequency-dependent rule model* and *(1iii) Threshold rule model*. We then *(2)* demonstrate the complex NBDA method under these examples, each time illustrating *(2i)* how the transmission rule can be specified, *(2ii)* how the data would be entered and the model fitted, *(2iii)* how the user can simulate data under this model (Figure 2, Supplementary Video 1), and *(2iv)* how the final output can be interpreted. Finally, we *(3)* directly test the ability of the Complex NBDA method to *(3i)* accurately quantify detect social transmission under complex contagions and *(3ii)* improve our estimates the influence of social connections on transmission in comparison to previous approaches.

*(1) Defining the complex NBDA model*

The standard NBDA model can be defined as follows:

$$\lambda_i(t) = \lambda_o(t)\left(s\sum_{j=1}^{N} a_{ij}z_j(t) + 1\right)(1 - z_i(t))$$

Where $\lambda_i(t)$ is the rate at which individual *i* acquires the target behaviour as a function of time, $\lambda_o(t)$ is a baseline rate function, *s* is a parameter determining the strength of social transmission, $a_{ij}$ is the network connection from *j* to *i*, and $z_i(t)$ is the 'status' of individual *i* at time *t*, (1= informed, has acquired the target behaviour; 0= naïve, has not acquired the target behaviour), and N is the number of individuals in the population. The rate at which an individual acquires the target behaviour by social transmission is proportional to $\sum_{j=1}^{N} a_{ij}z_j(t)$, the total connection to informed individuals at time *t*. *s*, therefore, gives the rate of transmission per unit connection relative to the rate of asocial learning of the target behaviour. The $(1 - z_i(t))$ term ensures that only naïve individuals acquire the behaviour, since NBDA is concerned with the time at which target behaviour is first acquired.

The complex NBDA model generalises this model as follows:

$$\lambda_i(t) = \lambda_o(t)(T(\boldsymbol{a_i}, \boldsymbol{z(t)}) + 1)(1 - z_i(t))$$

Where $\boldsymbol{a_i}$ is the vector of connections individual *i* has to all others in the network, $\boldsymbol{z(t)}$ is a vector giving the status of each individual in the network at time *t*, and $T(\boldsymbol{a_i}, \boldsymbol{z(t)})$ is a transmission function determining how the rate of transmission is determined by $a_i$ and $z(t)$. Therefore the standard NBDA is a special case of the complex NBDA with $T(a_i, z(t)) = \sum_j a_{ij} z_j(t)$. (Note also that Whalen & Hoppitt's (2016) expansion of NBDA is a special case where $T(a_i, z(t)) = f(\sum_j a_{ij} z_j(t))$.)

There are two variants of the standard NBDA model: time of acquisition diffusion analysis (TADA) and order of acquisition diffusion analysis (OADA). The former examines the times at which individuals acquired the target behaviour, whereas the latter takes only the rank order. TADA can have more statistical power, but requires assumptions to be made about the shape of the baseline rate function $\lambda_o(t)$. In contrast, OADA makes no assumptions about the shape of $\lambda_o(t)$, but instead makes the weaker assumption that it is the same for all individuals. In principle both OADA and TADA versions of the complex NBDA model can be fitted, but here we develop only the OADA variant. In the complexNBDA package, we extend the NBDA package (Hoppitt et al. 2020) to fit complex NBDA OADA models using maximum likelihood estimation in the R statistical environment (R Core Team 2020). In the examples below (and supported by the walk-throughs in *Supplementary Methods 1A-B & 2A-C*), we provide instructions on how to use this resource.

Within our package, we provide different social transmission rules (Figure 1) which we outline below. We then use the proportional rule (Figure 1B) to provide a detailed example of fitting, interpreting, and testing the model. We provide equivalent examples for the frequency dependent rule and the threshold rule in the supplementary information (*Supplementary Methods 1A-B*).

*(1i) Proportional rule model*

Here we fit a model where the rate of social transmission is proportional to the ratio of connections *i* has to informed individuals, i.e.

$$T(a_i, z(t)) = s \frac{\sum_j a_{ij} z_j(t)}{\sum_j a_{ij}}$$

where *s* gives the maximum rate of social transmission. Although the distinction between the sum of connections to informed individuals (i.e. simple contagion) and the proportion of connections to informed individuals (i.e. as expressed above) may appear subtle, conceptual work has demonstrated the difference may manifest as large changes in pathways of spread (*27*). Furthermore, such a transmission rule is well within the capacity of social species, and simply requires additional influence from individuals' uninformed (rather than just their informed) associates, or individuals copying the majority behaviour that they encounter. As such, it is a relatively unsophisticated strategy in comparison to the more advanced social learning behaviours animal species may show (*2, 6, 16, 22, 27*).

*(1ii) Frequency-dependent rule model*

The proportional model in example *1i* assumed that the rate of social transmission is proportional to the ratio of connections *i* has to informed individuals. It may be that individuals tend to show a conformity bias, being disproportionately more likely to copy what the majority of the population is doing i.e. a 'Frequency-dependent' rule. In the context of acquisition of novel behaviour, this bias might manifest itself as an individual being disproportionately more likely to acquire behaviour when that behaviour as being performed by the majority of the population. Evidence for this social learning strategy has been found for various species (*41-46*).

In the frequency-dependent transmission rule model, the transmission rule could be represented with the transmission function (adapted from McElreath et al. 2008):

$$T(\boldsymbol{a_i}, \boldsymbol{z(t)}) = s \frac{\left(\sum_j a_{ij} z_j(t)\right)^f}{\left(\sum_j a_{ij} z_j(t)\right)^f + \left(\sum_j a_{ij}\left(1 - z_j(t)\right)\right)^f}$$

Where the frequency dependence parameter f≥1, and s>0. When f=1 this model reduces to the proportional model above, and as f increases the strength of conformity bias increases. Consequently, if the 95% confidence intervals for f include only values >1 this can be taken as evidence for a conformity bias.

*(1iii) Threshold rule model*

Another example of a commonly considered social learning strategy is often referred to as the 'threshold' model. In this scenario, the probability of an individual adopting the behaviour remains low until their social connections to informed individuals surpasses a given threshold, and after this threshold further increases in informed associates have little influence on probability of adoption. While various lines of evidence exist for threshold-type strategies in human behaviour (*26*), this strategy is less researched in animal systems despite its potential importance (*40*).

In our example of the threshold model $T(\boldsymbol{a_i}, \boldsymbol{z(t)})$ is a modified logistic function:

$$T(\boldsymbol{a_i}, \boldsymbol{z(t)}) = \left(\frac{c}{1 - \frac{1}{1 + exp(ab)}}\right)\left(\frac{1}{1 + exp\left(-b(\sum_j a_{ij} z_j(t) - a)\right)} - \frac{1}{1 + exp(ab)}\right)$$

Where a,b,c >0. As with the standard NBDA, the rate of social transmission is zero when the total connections to informed individuals, $\sum_j a_{ij} z_j(t)$ =0. However, the rate of transmission increases suddenly as the threshold, *a*, is approached, to a maximum value of *c*. The parameter *b* determines how sharp the threshold effect is (see Figure 3). Our aim here is to generate a model with a clear sharp threshold, so we set *b*=3. However, an alternative is to estimate *b* as a parameter within the model.

### *(2) Fitting, generating and interpreting the model*

In this section, we use the proportional rule model (See 1i above and Figure 1B) to demonstrate how to *(2i)* specify the transmission rule, *(2ii)* enter the relevant data structure, *(2iii)* generate a simulated contagion under this rule and then *(2iv)* fit and interpret the model. We also provide equivalent examples for the frequency dependent rule (See *1ii* above) and the threshold rule (see *1iii* above) in *Supplementary Methods 1A-B*.

*(2i) Specifying the transmission rule*

To fit the proportional rule model, the user first needs to define a transmission function, which takes the arguments *par*: a vector of parameter values (in this case just a single value for *s* – see below), and *connectionsAndStatus*: a 2-row matrix- the first row giving the connections of *i* to other individuals in the network ($a_i$), the second giving their status ($z(t)$). We can then specify the proportional rule transmission function as:

```
proportionalRule<-function(par, connectionsAndStatus) {
  connectionToAll<-connectionsAndStatus[,1]
  statusOfOthers<-connectionsAndStatus[,2]
  totalConnectionToInformed<-sum(connectionToAll*statusOfOthers)
  totalConnection<-sum(connectionToAll)

  SocTransStrength<-par[1]
  if(totalConnection==0){
    #If there are no connections to anyone, social transmission rate=0
    rate<-0
  }else{
    rate<-SocTransStrength*totalConnectionToInformed/totalConnection
  }
}
```

*(2ii) Entering the data and fitting the model*

The user then creates a *complexNBDAdata* object using the same procedure as the regular NBDA method (e.g. as outlined in (*15*) or in this package description). The complexNBDAdata object contains the diffusion data in the form necessary to fit a complex NBDA model, using the function *complexNBDAdata* which takes as its arguments *label*: a character string labelling the data object; *assMatrix*: an association matrix or other social network, specified as a three dimensional array (see below); *orderAcq*: a vector giving the order in which individuals acquired the novel behaviour, given as number referring to the relevant row of the social network. To illustrate, imagine a 'toy' example with 5 individuals:

```
#Enter social network (or read in as a csv file)
socNet1<- rbind(c(0,0.5,0.5,0,0),
                c(0.5,0,0.5,0,0),
                c(0.5,0.5,0,0,0),
                c(0,0,0,0,0.8),
                c(0,0,0,0.8,0))
#This needs to converted into a three dimensional array
socNet1 <-array(socNet1,dim=c(dim(socNet1),1))
#Enter vector giving order of acquisition- number to match an individual's row in the network
orderAcq1<-c(4,5,2,3,1)
#Create complexNBDAdata object
toyData<-complexNBDAdata(label="toyData1", assMatrix= socNet1, orderAcq= orderAcq1)
```

An OADA model can be fitted using the complexOadaFit function, which takes as its arguments *complexNbdaData*: a complexNBDAdata object; *transmissionFunction*: a user specified transmission function; *noParTransFunct*: the number of parameters required by the transmission function (i.e. as outlined within each function) and *lower*: a vector of the lower bounds for each parameter (e.g. for the proportional rule model there is 1 parameter which is bounded above zero). A *startValue* vector can also be defined for the optimisation procedure (see *Supplementary Methods 2A-C* for full examples).

```
model_proportion<- complexOadaFit(complexNbdaData= toyData,
transmissionFunction= proportionalRule, noParTransFunct=1 , lower=0)
```

*(2iii) Simulating data generated by a specified model*

Given that this is a 'toy' example, with very few individuals we are unlikely to get a sensible output. Instead we can simulate a larger dataset, and fit a model to it in order to illustrate how to interpret the output. We can simulate data using the *simulateComplexDiffusion_OADA* function, which takes the arguments *par*: a vector of parameter values; *network*: a social network; *transmissionFunction*: a user-specified transmission function. For example:

```
#We will simulate a network with some structure but one can input a real life
network here.
#Set random number generator seed to get the same results as below:
set.seed(5)
simNet1<-matrix(runif(10000,0,1),100,100)
simNet1[simNet1<0.7]<-0
#here, simNet1 is a directional weighted association matrix, where the row for each
individual is incoming associations (e.g. the amount they observe each other
individual) and the column for each individual is their outgoing associations (e.g. the
amount they are observed by each other individual)

#Create some variation among individuals in connections (set here as max 3, but can
be anything)
multiplier<-matrix(runif(100,0,3),nrow=100,ncol=100)
simNet1<-simNet1*multiplier

#This needs to be in a 3D array by the requirements of the complexNBDA package
simNet1<-array(simNet1,dim=c(100,100,1))

#Simulate data
propModelSimData<-simulateComplexDiffusion_OADA(par=40, network=simNet1,
transmissionFunction = proportionalRule)
```

The object propModelSimData then provides the information on each individuals' time of adoption of the behaviour over the social network given the proportional contagion rule (see Figure 2 and Supplementary Video 1 for examples of simulated spread of each of the different behavioural contagions across the same simulated social network).

*(2iv) Interpreting model output*

```
#Fit a model to the simulated data, assuming the same transmission function
model_proportion<-complexOadaFit(propModelSimData, transmissionFunction = proportionalRule, noParTransFunct=1,startValue = 1,lower=0)
```

The maximum likelihood estimates of the parameters are contained in the slot model_threshold@mle, and standard errors in model_threshold@se. We can obtain a clear summary as follows:

```
#Display fitted parameters
data.frame(variable=model_proportion@varNames,MLE=model_proportion@outputPar,SE=model_proportion@se)
```

|   | variable | MLE | SE |
|---|---|---|---|
| 1 | 1 Transmission parameter 1 | 82.96274 | 116.2939 |

The standard errors give an idea of the precision of the estimates, but can be misleading in an NBDA, where there is often high precision for the lower bound of a parameter, but low precision for the upper bound (see *Supplementary Methods 2A-C*). Furthermore, standard errors cannot always be derived. Therefore, we recommend that researchers use the profile likelihood technique to derive 95% confidence intervals for parameters (see *Supplementary Methods 2A* "Proportional rule example.R", for how to do this in the *complexNBDA* package). In this case the 95% confidence interval is estimated to be 3.4 – 2099. So, we can see that we can set a clear lower plausible limit on the strength of social transmission and can clearly exclude s=0, but have little information about the plausible upper limit (hence the high standard error). In general, we suggest that instead of interpreting the exact point estimates for parameters in a complex NBDA, researchers obtain 95% confidence intervals and interpret the biological significance of these, especially in cases where the profile -log-likelihood is highly asymmetrical. So, in this example, the key conclusion would be that "the maximum effect of social transmission (*s*) is at least 3.4x the rate of asocial learning" since we do not have much information about the upper plausible limit.

We can obtain the Akaike's Information Criterion, corrected for sample size (AICc) as follows:

```
model_proportion @aicc
```

Getting a value of 722.81. This can be compared to other models with different transmission functions to see which best predicts the order of acquisition. For example, we can fit a standard NBDA as follows:

```
#Create a standard nbdaData object
nbdaData1<-nbdaData(label="propStandard",assMatrix=simNet1,orderAcq = propModelSimData @orderAcq)
#Fit the model to the data
model_ standard <-oadaFit(nbdaData1)
#Get the AICc
```

```
model_standard@aicc
```

Giving a value of 728.53- therefore with a difference of 5.72 in favour of the proportional rule model, it is clearly supported over the standard NBDA in this case- as we would expect since that is the model used to generate the data. We can also compare to a model with purely asocial learning as follows:

```
model_asocial<-oadaFit(nbdaData1,type="asocial")
model_asocial @aicc
```

We get a value of 727.48, a difference of 4.67 in favour of the proportional rule model.

For the full R script for stage *2i* to *2iv*, please see *Supplementary Methods 2A*. Further, for equivalent examples for Sections *2i-2iv* of the other transmission rules (frequency dependent and threshold), please see *Supplementary Methods 1A-B & 2B-C.*

**(3) Testing model performance**
In this section, we present the results of data simulations under each social transmission rule (Figure 2, Supplementary Video 1), and test how successful the complex NBDA method is at recovering the parameter values used to generate the data, and how much better it performs than standard NBDA. We also show how this information can be used to adjust the analysis to ensure the validity of inferences. We anticipate that researchers will often wish to use their own transmission rules, and even if they use the transmission rules presented here, performance will depend on sample size and network structure. Consequently, we recommend that researchers conduct equivalent simulations to test and adjust model performance for their analyses. Code for running the simulations is provided in the script files for each of the examples.

*(3i - ii) Proportional rule and frequency dependent models*

Since the proportional rule simulations (Figure 2B) are a special case of the frequency dependent model (with $f$=1) (Figure 2C), we run a single set of simulations for these two transmission rules (code provided in the *Supplementary Methods 2B* "Frequency dependent example.R"). In each simulation we generated a random network of 100 individuals in which the total connection to other individuals had a mean of approximately 38 and standard deviation of approximately 23 (*Supplementary Methods 2B*). We simulated a diffusion through the entire population (Figure 2C), following the frequency dependent model. We then fitted two OADA models to the simulated data assuming frequency dependent transmission and the proportional rule, and recorded the estimated parameter values, AICc, and whether the true values of $s$ and $f$ where within their 95% confidence intervals. Finally, we fitted a standard OADA model to the data, and a model of asocial learning and recorded the AICc in each case. We varied the value of the maximum social transmission or 'size' parameter, $s = \{0,5,10,30\}$, and the frequency dependence parameter, $f= \{1,2,3,5\}$. 1000 simulations were run for each combination of $s$ and $f$, except for s=0 (asocial learning only), where the value of $f$ can make no difference, so only a single set of 1000 simulations was run.

We found that the asocial model was correctly favoured by AICc 82.5% of the time when $s$=0. When $f$=1, the proportional rule model was favoured over the frequency dependent model, and was more likely to be correctly favoured over the asocial model and standard NBDA as $s$ increased. When f≥2 the frequency dependent model became increasing likely to be correctly favoured over the other three models as $f$ and $s$ increased (see Figure 4).

The exact numbers seen in Figure 4 are not general results for these parameter values, since the power to detect social transmission will depend on sample size and network structure. However, the results show that the complex NBDA method is capable of distinguishing the frequency dependent model, proportional rule model, asocial learning and the standard NBDA model, and that power to do so will increase as the strength of social learning ($s$) increases and as the strength of frequency dependence ($f$) increases.

We suggested above that, in general, instead of interpreting the exact point estimates for parameters in a complex NBDA, researchers obtain 95% confidence intervals and interpret the biological significance of these. Consequently, we wished to know what proportion of cases the true values lay within the 95% CIs ('coverage', ideally this should be close to 95%). Here we examine only the frequency dependent model, performance of the proportional rule model is examined in *Supplementary Methods 2A* "Proportional rule Example.R".

For the frequency dependence parameter $f$, we find coverage is sufficiently close to 95% (see Table S1). This suggests that the 95% C.I can be trusted as providing a plausible range for $f$, at least for the network and sample size in our simulated datasets. In contrast, for the size parameter $s$ the coverage is well short of 95% in all cases (see Table S2). This suggests the 95% confidence intervals are misleadingly narrow. Consequently, in the *Supplementary Methods 2B* we show how to adjust the 95% C.I. to ensure that it contains 0 in approximately 95% as required. When this technique is applied to the dataset from Methods Section 2, we find the 95% C.I. for $s$ is widened from 10.9 – 207 to 6.6 – 542.

*(3iii) Threshold Model*

In each simulation we generated a random network of 100 individuals in which the total connection to other individuals had a mean of approximately 25 and standard deviation of approximately 4 (*Supplementary Methods 2C* "Threshold function example.R"). We simulated a diffusion through the entire population (Figure 2D), following the threshold model (See *1iii* and Figure 1C). We then fitted an OADA model to the simulated data assuming the threshold transmission rule, and recorded the estimated parameter values, AICc, and whether the true values of $a$ and $c$ where within the 95% confidence intervals. Finally, we fitted a standard OADA model to the data, and a model of asocial learning and recorded the AICc in each case. We varied the value of the threshold parameter, $a$ = {2.5,5,10,15}, and the max social transmission or 'size' parameter, c= {0,5,10}. 1000 simulations were run for each combination of $a$ and $c$, except for c=0 (asocial learning only), where the value of $a$ can make no difference, so only a single set of 1000 simulations was run.

We found that the asocial model was correctly favoured by AICc 85.1% of the time when $c$=0. When c>0 the threshold model tended to be correctly favoured over the asocial model and the standard NBDA model. As both a and c increased, the threshold model became increasing likely to be favoured by AICc (see Figure 4).

As before, the exact numbers seen in Figure 5 are not general results for these parameter values, since the power to detect social transmission will depend on sample size and network structure. However, the results show that the Complex NBDA method is capable of distinguishing the threshold model from both asocial learning and the standard NBDA model, and that power to do so will increase as the strength of social learning (c) increases and as the threshold (a) moves away from zero.
For the location parameter *a*, we find coverage is reasonably close to 95% except when the real value of *c* is close to zero, when the 95% confidence intervals appear to be slightly too narrow (see Table S3).

This suggests that if the 95% C.I. is clearly a long way from zero it can be trusted as providing a plausible range for *c*, but if it contains values fairly close to zero it may need adjusting. We describe this process in *Supplementary Methods 2C*. When this technique is applied to the dataset from the *Supplementary Methods 2C* 'Threshold function Example.R', we find the 95% C.I. for *f* is widened slightly from 4.8 – 6.2 to 4.6 – 6.3. For the size parameter *c* the coverage is close to 95% in all cases except for *c*=0 (see Table S4). This suggests the 95% confidence intervals might, in certain circumstances, give an imprecise picture of the level of confidence against the null hypothesis of no social transmission. Consequently, *Supplementary Methods 2C*, we also show how to adjust the lower end of the 95% C.I. to ensure that it contains 0 in 95% of cases when *c*=0, as required. Again, when this technique is applied to the dataset from *Supplementary Methods 2C*, we find the 95% C.I. for *c* is widened from 4.4 – 126 to 3.1 – 126. As such, as these findings suggest that the coverage of C.I.s can potentially differ depending on datasets (e.g. sample sizes, network structure), we recommend that researchers conduct equivalent simulations (as shown above) to test and adjust model performance for their analyses (using the code provided in the script files for each of the examples as a framework for this).

**DISCUSSION**
We provide a new method for assessing the social spread of behaviours in networks, and demonstrate that this approach provides new insight into the mode and pathways of behavioural transmission, and addresses limitations of previous approaches (*14, 27*). Specifically, given the extent of emerging empirical evidence that behaviours often spread as complex contagions (*26, 27*), our method facilitates the integration of social transmission rules (i.e. social learning strategies) directly into the framework most frequently used for understanding behavioural spread in animal social systems (NBDA), providing a flexible tool for calculating how social bonds and strategies influence behavioural contagion across social networks. Below, we (i) outline how this method and the assessment of results conceptually relates to current understanding of sociality in animal systems, (ii) discuss the implications for behaviour and ecology and the various potential applications, and finally (iii) highlight the primary opportunities of further work that are now possible given the current foundations.

*Animal Sociality and Understanding Complex Contagions*

Animal social learning is a topic that is extensively investigated and conceptually explored in depth (*2, 47*). Particular attention has been paid to the different strategies that individuals may employ when adopting socially-informed behaviours, such as disproportionately conforming with the majority of individuals (*47*), or copying particular individuals or learning only under certain contexts. As these strategies can shift the fitness benefits of social information use (*2, 45*) and thus shape the evolution of learning, it is perhaps unsurprising that evidence for various forms of social learning strategies is widespread. Despite the widespread recognition of the importance of social learning for the individual, the population-level consequences of these strategies remain relatively unexplored. Importantly, under a social learning strategy, the diffusion of the behaviour across the population is predicted to follow a spreading pattern synonymous to the complex contagions commonly considered for human studies, rather than depending simply on the raw number of connections among individuals (*27*). Yet, due to the absence of a general analytical framework for investigating how social learning strategies of individuals shape the behavioural contagion dynamics across the population, empirical research until now has largely relied on developing bespoke analytical approaches customised to each study (*40, 48, 49*). As such, we hope that generality and suitability of Complex NBDA for applications across systems will support new research in this field, particularly in terms of assessing the evidence for, and mechanics behind, complex contagions in animal systems.

One of the most prominent social learning strategies considered in animal populations is conformity (*42, 45, 46*), which has generated much interest due to its potential for maintaining group-typical behaviours and animal culture (*41, 43, 44*). Contemporary assessments of patterns of behavioural diffusions have suggested conformity may be in operation as new behaviours spread across natural social networks (*41-43*). However, "conformity" in these studies was primarily based on emergent population-level patterns, rather than on individual-level decisions given local social encounters, leading to discussion around these conclusions (*50-55*). Our new complex NBDA method presented here provides a direct opportunity to address this, whereby the researchers can clearly examine whether a simple diffusion best explains the behavioural spread, or whether model fit is significantly improved by considering a frequency-dependent complex contagion in which individuals express a 'disproportionately conform with the majority' strategy.

Another commonly considered class of the many different social learning strategies (other than conformity) that individuals can potentially express is 'model bias', whereby individuals bias their copying towards a particular type of individual/model (*19, 47, 56*). Interestingly, recent NBDA approaches have proven useful for examining model biases, and have – for instance - found evidence of wild birds copying those of the same species (*57*) and those that were experimentally designated as 'relevant tutors' to them (*19*), and for primates learning from those of high prestige/status (*58, 59*). More specifically, such insights were enabled by a multi-matrix approach to NBDA, where separate social networks within separate NBDA models are fitted as the predictor of the observed diffusion, and conclusions are then based on which model (and associated social network) provides the best fit (*14*). Nevertheless, examining who 'negatively' influences individuals (in terms of reducing the probability of them expressing a new behaviour), and investigating when individuals will learn (given their other connections), has remained challenging. Indeed, even within the multi-matrix approaches, the models remained essentially dyadic, wherein each dyadic connection was treated as independent from other dyadic connections. By addressing these gaps, our method (which can be easily combined with a multi-matrix approach) will contribute substantially to furthering understanding of how model based social learning shapes the spread of new behaviours at the population-level.

We also hope that our method will indicate when the distinction between simple contagions, complex contagions, and various social learning rules governing complex contagions, are particularly important. For instance, use of simple diffusion models may be responsible for potentially under-estimating the influence of social connections on the adoption of behaviours in fish (*60*) and otters (*61*). Going forward, it will help to consistently determine how social learning rules employed by individuals shape the spread of behaviours across populations, and quantifying how this governs the strength and rate of behavioural contagions.

*Implications and Potential Applications Across Topics*

Although our new method is directly useful for understanding animal social learning and social networks, this new approach for considering behavioural spread (and associated tools) also has wider implications wider implications across topics and presents a diverse range of further opportunities and applications. Firstly, our understanding of social learning can determine how we expect populations to respond to ecological change. The spread of knowledge regarding the environment is governed by the social connections within and across generations ("horizontal" transmission and 'vertical' non-genetic inheritance, respectively) (*7, 62, 63*). Therefore, through allowing the consideration (and simulation) of complex contagions of behaviour, the approach we present holds implications for this important topic within ecology.

The framework provided here may also inform work on the evolution of networks in animal systems. Specifically, recent studies have generally conceptualised the 'optimality' of the network as its efficiency for social transmission (*64, 65*). But, these considerations have been based on only considering the efficiency of simple contagions, despite that complex contagions may vary dramatically from these and play a major role in the spread of animal behaviours (*27*). Thus, our method will help to expand the notion of optimality of social networks in animal systems, prompting the consideration of structures' efficiency for the spread of behaviours under different social learning rules. This expansion may provide new insights into how animal social network structure may evolve, and how selection may act on these social systems.

The study of social networks and the social spread of behaviour has potentially important applications for animal conservation (*63, 66*). Our approach is directly related to two primary aspects of these applications, namely (i) properly understanding how harmful (or helpful) behaviours spread socially in conservation-relevant populations (*63, 67*), and (ii) addressing how important socially learnt behaviours (such as mammalian migration strategies (*68*)) are maintained across generations (*66*). Furthermore, recognising the potential for complex contagions, and appropriately analysing them, also holds applications for conservation science in terms of understanding anthropogenic influences. Understanding how new conservation initiatives, conservation-positive attitudes, or conservation behaviours such as bycatch reduction strategies (*69*), may spread through human societies, is of much importance, but research in this area often considers such processes as simple contagions (*70*), and it has recently been acknowledged that new approaches are needed for considering human social networks in relation to conservation outcomes (*69, 71*). As our method can also be applied to human social networks, this provides the opportunity to assess the significance of complex contagions in relation to the spread of conservation-related behaviours.

Developing techniques that allow a fine-grained approach to understanding contagion within social networks is also likely to have applications for understanding the dynamics of other types of contagion in animal systems, particularly disease (*72*). Understanding disease spread in natural animal populations is tied to a range of challenges (*72, 73*), such as incomplete sampling of individuals and social interactions between them (*73, 74*), and restricted knowledge of the pathogen in terms what the relevant contacts are and which kind of social dynamics promote transmission (*72, 73, 75*). As such, quantifying the social component of disease spread, and comparing evidence for competing hypotheses in terms of contact types/patterns that promote transmission, is difficult, and generalised tools to carry out such analyses are lacking (*75*). Techniques such as those that we develop here will not only be useful for building a general understanding of how complex social dynamics contribute to disease contagion, but also for determining how simple disease contagions differ from the spread of behaviours. Through identifying how the pathways and mechanism of contagion differ between diseases and behaviours, this will insight into ways that disease transmission can be controlled without also drastically reducing the potential for the spread of useful information.

*Further Opportunities*

The method described here also provides a platform for additional features to be built in the future. For instance, while the framework already allows users to specify their own social learning strategies (or transmission rules), test for their operation in real populations, and investigate their impact via simulations (Figure 2, Supplementary Video 1), a further advance would be to consider that social learning strategies may vary between individuals, or time-frames. Indeed, evidence exists that individuals may vary in the threshold at which they adopt behaviours (*28*) or differ entirely in the transmission rules they employ (*56*).

A substantial development of this method would allow for feedback between behavioural adoption and social network connections, which can yield variable outcomes depending on the contagion process underway (*27*). Current research has focused heavily on how individuals' network connections may influence their propensity to gain information, yet adopting a socially-informed behaviour may change an individual's network position (e.g. Figure 6). For instance, learning the location of a new food patch or foraging technique may alter an individual's behaviour directly (e.g. by allowing them to associate with more individuals) or alter the propensity of other individuals to interact with them (e.g. as others attempt to gain the information by associating with them) (*35*). Alternatively, if the adoption and transmission of behaviour is highly dependent on clustered interactions within tight cliques of individuals (i.e. in certain complex contagions), then feedback can potentially reduce behavioural spread (*27*). Although the feedback between information flow, behavioural change, and social network positions is currently under-explored, a similar feedback process is known to be important within disease, whereby contracting an infectious disease may alter either an individual's own social behaviour or influence how others interact with them (*73, 76*). These changes to social interaction patterns can thus alter transmission dynamics across the network (*77, 78*).

Just as the feedback between social connections and behavioural adoption may influence contagion patterns, interactions between multiple different contagion processes can themselves govern spreading dynamics (*26, 33*). Research within human systems has found evidence for interacting contagions in terms of behaviours interacting with other behaviours (*79*) as well as behaviours and disease interacting with one another (*29, 80*). Nevertheless, generalised methodologies for investigating how contagions interact and shape one another's subsequent spread in real-world settings do not exist. As such, another significant future possibility for the method we propose here would be to integrate multiple contagions, and simultaneously quantify their social components and the interaction between them.

*Conclusion*
Our framework allows for the consideration, quantification, and simulation of behavioural contagions, and it holds various opportunities for advancing understanding of the ecology and evolution of social spread within natural systems. However, several future developments of the current method, as highlighted above, are also conceptually possible and may lead to further advances for research in this area. Therefore, we hope that the concepts and tools provided here, and the description of their usage and applications, will enable researchers to investigate the pathways and modes of spread within their own study systems under a common framework, and ultimately provide generalised insight into transmission processes in animal populations.

**Acknowledgements**
JAF was supported by a research fellowship from Merton College and BBSRC (BB/S009752/1). JAF, KBB & BCS acknowledge funding support from NERC (NE/S010335/1).

**Author contributions**
JAF and WH conceived the study, designed the complexNBDA framework, carried out the analysis, wrote the first draft of the manuscript and built the figures and supplementary movie. All other authors provided insight into the analyses, contributed to discussions, and provided feedback on the manuscript.

**Conflict of interests**
None of the authors report a conflict of interest.

**Code and data accessibility**
All of the resources (code and generated data) are available at https://github.com/whoppitt/complexNBDA and the scripts are also included in the supplementary information (See Supplementary Methods 2).

MAIN TEXT FIGURES AND LEGENDS

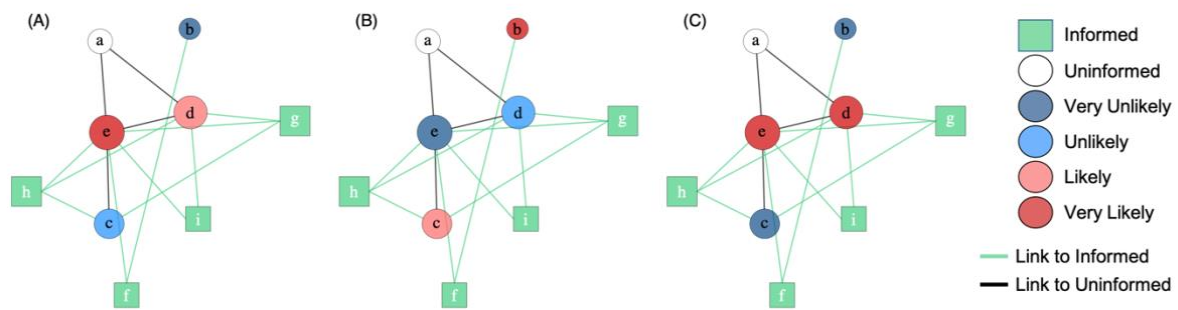

**Figure 1.** Illustrative example of how different behavioural contagion scenarios can result in different relative likelihoods of the same individuals adopting the behaviour, even under identical social network structures. Across all panels an identical network structure is maintained, and the node size shows the number of social ties each individual has, while node shape shows individuals' state: circles denote uninformed individuals (i.e. individuals *a*, *b*, *c*, *d* and *e* haven't yet adopted the behaviour) and the green squares represent 'informed' individuals (i.e. individuals *f*, *g*, *h*, and *i* have already adopted the behaviour). The adjoining lines show the social ties between individuals, with links to informed individuals as green lines and links between uninformed individuals as black lines. The colour of the circles represent the relative likelihood of each uninformed individual's chance of being the next individual to adopt the behaviour given their current social links, and given the contagion rule at play within each panel. Individual '*a*' has no colour across all panels as it has no links to informed individuals so cannot be the next to adopt the behaviour. **(A)** Simple contagion; adopting the behaviour is dependent on the sum of social links to informed individuals, so individual '*b*' is the least likely ('very unlikely') to adopt the behaviour (just 1 link to an informed individual), followed by '*c*' (2 links to informed – 'unlikely'), then '*d*' (3 links to informed – 'likely') and '*e*' has 4 links to informed individuals so is relatively 'very likely'. **(B)** Proportional rule; adopting the behaviour is dependent on the proportion of social links to informed individuals over uninformed individuals, so the pattern is reversed from the simple contagion scenario, and now individual '*b*' is 'very likely' to adopt the behaviour (100% of its links are to informed individuals), '*c*' is 'likely' to be the next to adopt the behaviour (66.7% to informed, i.e. 2 links to informed, 1 to uninformed), '*d*' is less likely still (3 links to informed, 2 to uninformed), and now '*e*' is the least likely to adopt the behaviour despite having the most links to informed individuals as it also has the most links to uninformed individuals (4:3). **(C)** Threshold rule; adopting the behaviour is dependent on having a threshold amount of links to informed individuals (here set as 3 links) where those that have fewer than this are very unlikely to adopt the behaviour (individuals '*b*' and '*c*') while those that meet this threshold are relatively very likely (individuals '*d*' and '*e*') to be the next to adopt the behaviour. Therefore, across the panels, these examples illustrate how the contagion rule at play can significantly alter the order in which uniformed individuals adopt the behaviour, which informed individuals are successful in transmitting it, and the routes of social spread across the network.

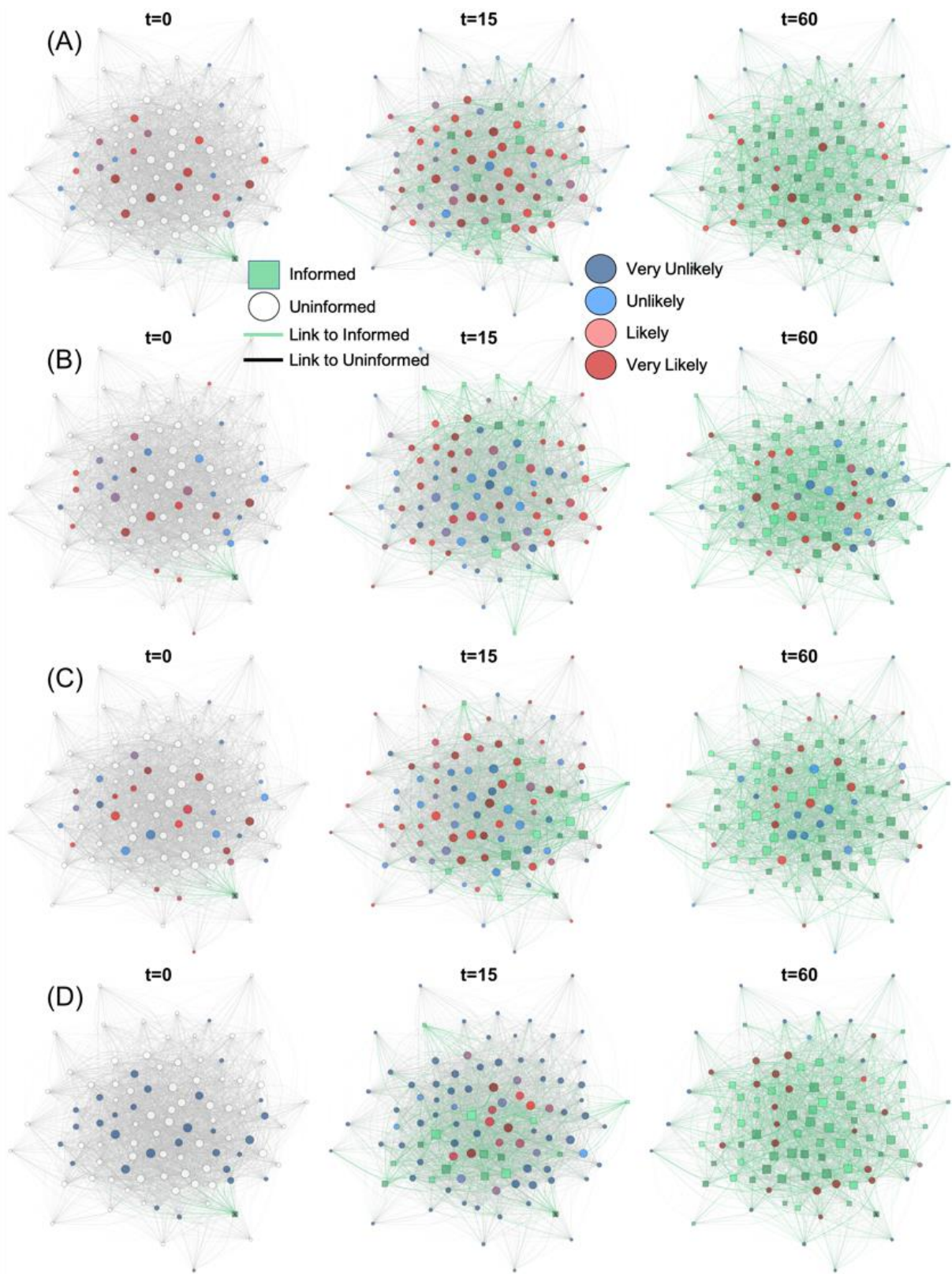

**Figure 2.** Examples of the different behavioural contagion scenarios simulated across the same simulated social network. The rows show each scenario **(A)** Simple contagion **(B)** Proportional rule **(C)** Frequency Dependent rule, and **(D)** Threshold rule. The columns show the time-step within the contagion (ranked times of number of individuals adopted). Across all panels an identical network structure is maintained, and each has the same start node. The node size shows the number of social ties each individual has, while node shape shows individuals' state: circles denote uninformed individuals and the green squares represent 'informed' individuals, with the darkness of the green showing the time since adopting the behaviour (darker = earlier adoption). The adjoining lines show the social ties between individuals, with links to informed individuals as green lines and links between uninformed individuals as grey lines. The colour of the circles (uninformed nodes) represent the relative likelihood of each uninformed individual's chance of being the next individual to adopt the behaviour given their social links and given the contagion rule at play within each panel. White nodes show individuals with no links to informed individuals at that time step. See Supplementary Video 1 for animated version (contagion through the time-steps) of the figure.

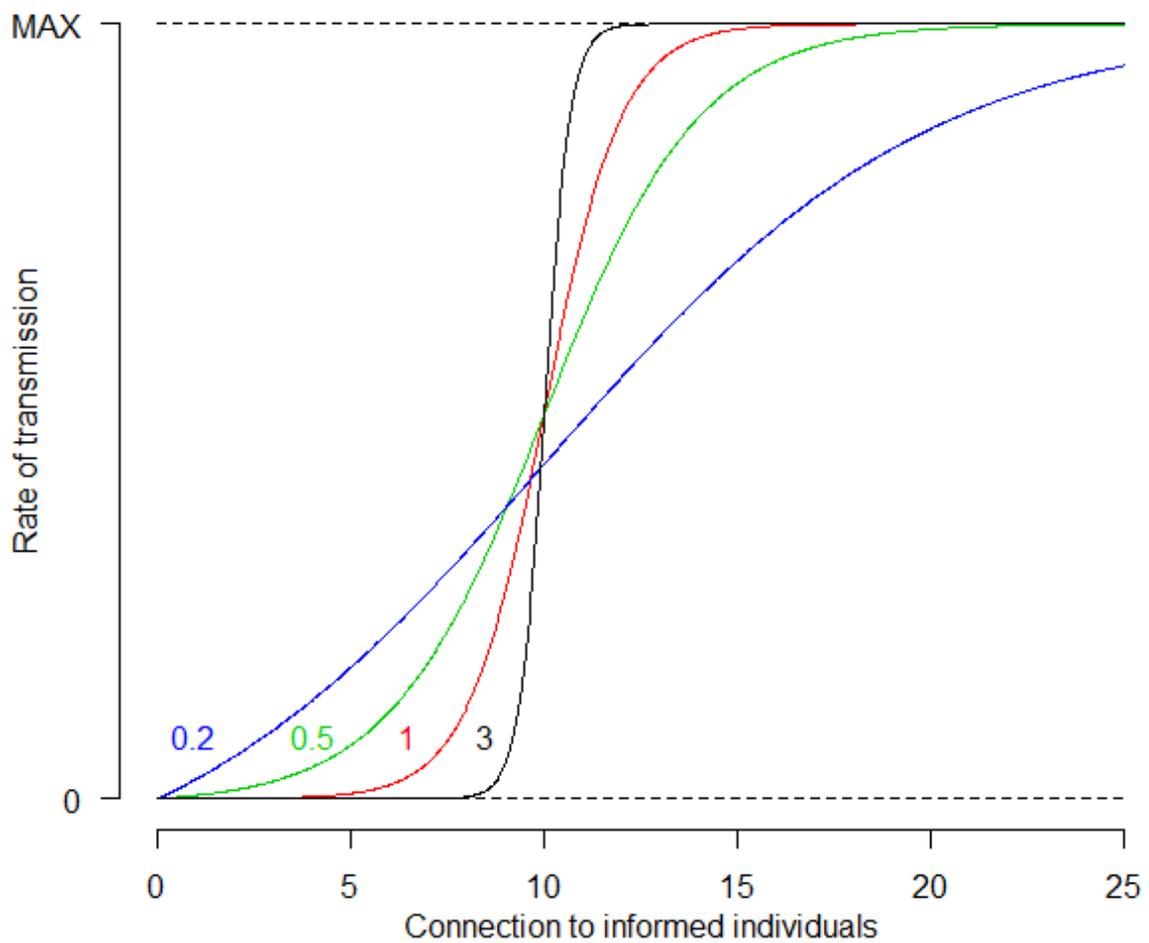

**Figure 3.** The rate of social transmission as a function of connection to informed individuals under the threshold rule model with different values of b (a=10, c=10). The higher the value of b, the sharper the threshold. When using the threshold model within the following examples, we set b=3.

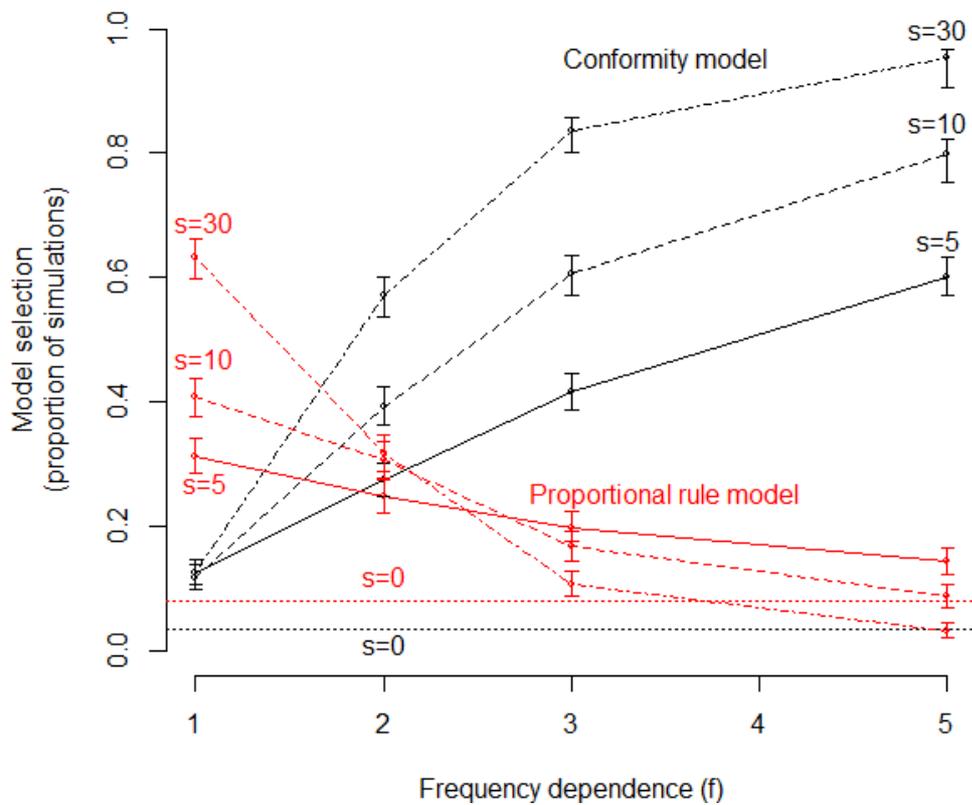

**Figure 4.** The proportion of frequency-dependent behavioural contagions simulations in which the proportional rule model (red lines) and the conformity (i.e. frequency dependent) model (black lines) were favoured (in a four way comparison including asocial learning and the standard NBDA model), as a function of the value of the frequency dependence (conformity bias) parameter, *f*, and maximum social transmission strength parameter, *s*. Results for *s*=0 are shown as a dotted horizontal line since the value of *f* makes no difference when *s*=0. Error bars show 95% confidence intervals for each point.

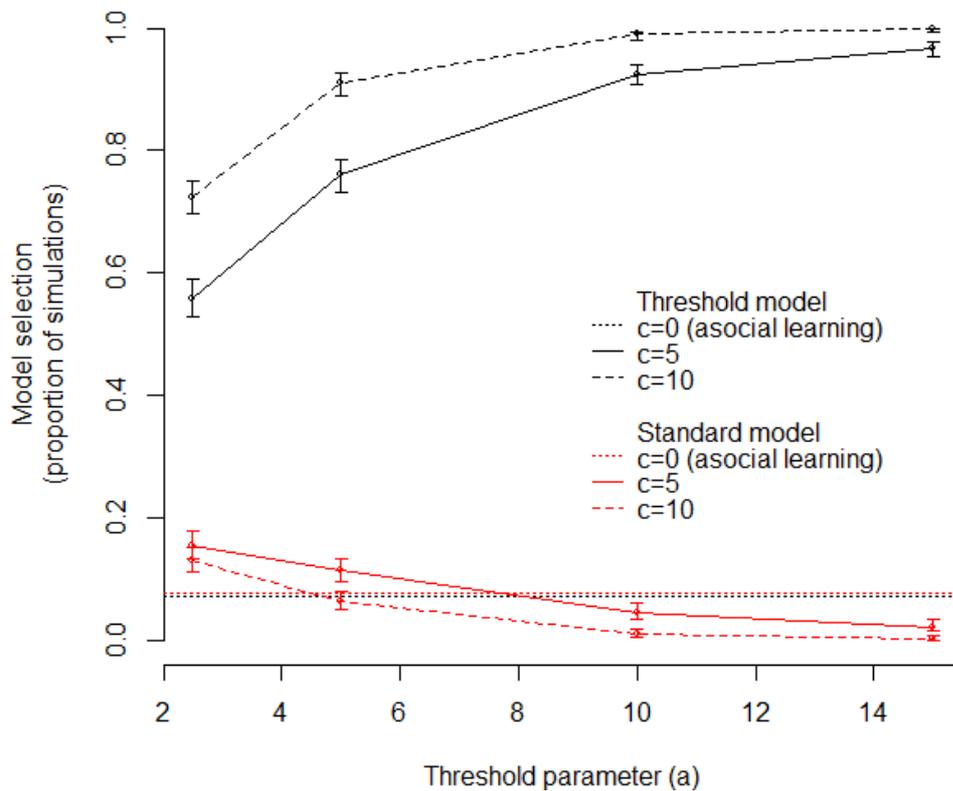

**Figure 5.** The proportion of threshold behavioural contagion simulations in which the threshold model and standard (i.e. simple contagion) model were favoured (in a three way comparison including asocial learning), as a function of the value of the threshold parameter, a, and maximum social transmission strength parameter, c. Results for c=0 are shown as a dotted horizontal line since the value of a makes no difference when c=0. Error bars show 95% confidence intervals for each point.

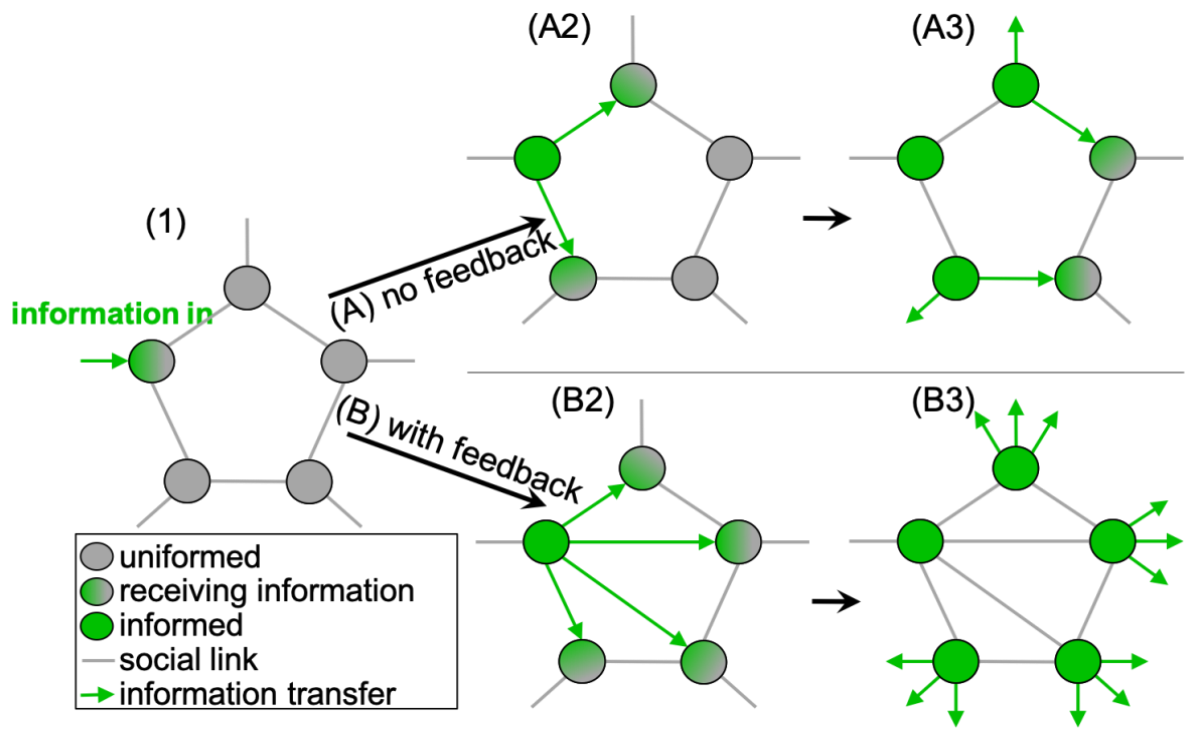

**Figure 6.** Simple example of Information flow and behaviour change through a toy network subsection. Individuals (circles) transfer information through social associations (lines) over sequential stages (1, 2 & 3). Scenario (A) Individuals do not change their associations in response to information flow (i.e. no feedback occurs). Scenario (B) Individuals gain social links upon acquiring information and adopting the behaviour (i.e. feedback occurs).

SUPPLEMENTARY INFORMATION:

# Analysing the Social Spread of Behaviour: Integrating Complex Contagions into Network Based Diffusions

Josh A. Firth*[†,1,2], Gregory F. Albery[3,4], Kristina B. Beck[1,5], Ivan Jarić[6,7], Lewis G. Spurgin[8], Ben C. Sheldon[1] & Will Hoppitt*[†9]

[1]Department of Zoology, Oxford University, Oxford, UK

[2]Merton College, Merton Street, Oxford, UK

[3]Department of Biology, Georgetown University, Washington, DC

[4]Institute of Evolutionary Biology, University of Edinburgh, Edinburgh, UK

[5]Department of Behavioural Ecology and Evolutionary Genetics, Max Planck Institute for Ornithology, Seewiesen, Germany.

[6]Biology Centre of the Czech Academy of Sciences, Institute of Hydrobiology, České Budějovice, Czech Republic,

[7]University of South Bohemia, Faculty of Science, Department of Ecosystem Biology, České Budějovice, Czech Republic

[8]School of Biological Sciences, University of East Anglia, Norwich, UK

[9]School of Biological Sciences, Royal Holloway, University of London, Egham, UK

*Correspondence: Joshua.firth@zoo.ox.ac.uk & whoppitt@mac.com [†]Joint lead authors

**SUPPLEMENTARY INFORMATION CONTENT:**

**(1) Supplementary Methods 1: Examples of Social Transmission Rules**

    -1A Frequency-dependent Rule

    -1B Threshold Rule

**(2) Supplementary Methods 2: Description of Supplementary R Scripts**

**Available at: https://github.com/whoppitt/complexNBDA**

    -2A-C Examples for building, simulating and testing each social transmission rule

    -2D-H Base scripts for the complexNBDA package

**(3) Supplementary Tables**

    -3A-B Additional Information on Performance Testing for Frequency-dependence Model

    -3C-D Additional Information on Performance Testing for Threshold Model

**(4) Supplementary Video 1 Description**

## SUPPLEMENTARY METHODS

### Supplementary Methods 1: Examples of social transmission rules

The primary manuscript uses the proportional rule (Figure 1B; Figure 2B) as an example for specifying, fitting, simulating and testing the model (Methods Section 2). Here, we provide the equivalent examples but for social transmission under the Frequency-dependent rule (Methods Section 1ii; Figure 2C) and under the threshold rule (Methods Section 1iii; Figure 1C; Figure 2B). All code is available at https://github.com/whoppitt/complexNBDA.

*Fitting, generating and interpreting the models*

### Supplementary Methods (1A) Frequency-dependent Model
The Frequency-dependent transmission function can be coded in R as follows:

```r
frequencyDependentRule<-function(par, connectionsAndStatus){
  connectionToAll<-connectionsAndStatus[,1]
  statusOfOthers<-connectionsAndStatus[,2]
  totalConnectionToInformed<-sum(connectionToAll*statusOfOthers)
  totalConnection<-sum(connectionToAll)

  SocTransStrength<-par[1]
  f<-par[2]
  if(totalConnection==0){
    #If there are no connections to anyone, social transmission rate=0
    rate<-0
  }else{
    rate<- sum(
      SocTransStrength*sum(connectionToAll*statusOfOthers)^f/
        (sum(connectionToAll*statusOfOthers)^f+sum(connectionToAll*(1-statusOfOthers))^f)
    )
  }
  return(rate)
}
```

We can simulate data for s=30, f=4, and fit a model to the simulated data as follows:

```r
#We will simulate a network with some structure but one can input a real life network here
#Set random number generator seed so we get the same results
set.seed(7)
simNet1<-matrix(runif(10000,0,1),100,100)
simNet1[simNet1<0.7]<-0
#Create some variation among individuals in incoming connections
multiplier<-matrix(runif(100,0,3),nrow=100,ncol=100)
simNet1<-simNet1*multiplier

#This needs to be in a 3D array by the requirements of the complexNBDA package
```

```
simNet1<-array(simNet1,dim=c(100,100,1))

#Simulate data
freqDepModelSimData<-simulateComplexDiffusion_OADA(par=c(30,4), network=simNet1,
transmissionFunction = frequencyDependentRule)
```

The object freqDepModelSimData then provides the information on each individuals' time of adoption of the behaviour over the social network given the Frequency-dependent contagion rule (see Figure 2C for example).

Now we can fit the model (see Supplementary Material example script "Frequency-dependent example.R" for recommendations on choosing *startValue*):

```
freqDepModel<-complexOadaFit(freqDepModelSimData, transmissionFunction =
frequencyDependentRule, noParTransFunct=2,startValue = c(5,5),lower=c(0,0.2))

data.frame(variable= freqDepModel@varNames, MLE= freqDepModel@outputPar, SE=
freqDepModel@se)
```

Obtaining:
```
                  variable       MLE         SE
1 1 Transmission parameter 1 62.180328 55.2227509
2 2 Transmission parameter 2  4.232584  0.9873376
```

In this case f has been estimated close to its true value of 4, whereas s has been overestimated at 62.2. Code for obtaining 95% confidence intervals is provided in the Supplementary Material example script "Frequency-dependent example.R". These were found to be 2.4 – 6.5 for *f* and 10.9 – 207 for *s*. Since *f*=0 is a fair way outside the 95% confidence interval for *f*, this would be taken as strong evidence of a conformity bias.

We can obtain the AICc for the model:

```
freqDepModel@aicc
```

Obtaining 713.08. We can then compare this to a proportional rule model (f=1), standard NBDA model and asocial model:

```
model_proportion<-complexOadaFit(freqDepModelSimData, transmissionFunction =
proportionalRule, noParTransFunct=1, lower=0)
model_proportion@aicc
# [1] 718.7073
#Create a standard nbdaData object
nbdaData1<-nbdaData(label="propStandard",assMatrix=simNet1,orderAcq =
freqDepModelSimData @orderAcq)
#Fit the model to the data
model_standard <-oadaFit(nbdaData1)
#Get the AICc
model_standard@aicc
# [1] 727.6171

model_asocial<-oadaFit(nbdaData1,type="asocial")
```

```
model_asocial@aicc
# [1] 727.4788
```

Finding that the Frequency-dependent model has reasonably more support than the proportional model (difference in AICc= 5.6) and far more support than the standard NBDA (14.5) and asocial model (14.4).

**Supplementary Methods (1B) Threshold model**
To fit the threshold model, the user first needs to define a transmission function, which takes the arguments *par*: a vector of parameter values (a, c), and *connectionToInformed*: the elementwise vector $a_{ij}z_j(t)$ (the connections to informed individuals are present, but the connections to naïve individuals are replaced with zero).

```
logisticLearning<-function(par, connectionToInformed){
  thresholdSharpness<-3
  thresholdLocation<-par[1];
  size<-par[2];
#The first parameter gives the location of the threshold
#The second parameter gives the (maximum) strength of social transmission
  rate<-size/(1 - 1/(1+exp(thresholdSharpness*thresholdLocation)))*
    (1/(1+exp(-thresholdSharpness*(sum(connectionToInformed)-thresholdLocation))) -
1/(1+exp(thresholdSharpness*thresholdLocation)))
  rate
}
```

Note that this transmission function could be defined using the *connectionsAndStatus* argument. However, defining a transmission function using *connectionToInformed* results in faster computations when fitting the model, so if a transmission function can be specified using *connectionToInformed* then this is to be preferred.

We will now simulate data from the threshold model and fit an OADA model to the simulated data:

```
#Set the seed for the random number generator, so the same results are obtained as here
set.seed(5)
#Generate a social network of 100 individuals with some structure (though one can
alternatively input a real life network)
simNet1<-matrix(runif(10000,0,1),100,100)
simNet1[simNet1<0.7]<-0
simNet1<-array(simNet1,dim=c(dim(simNet1),1))

#Simulate data assuming the logisticLearning transmission function, with parameters a= 5,
c=10
thresholdModelSimData<-simulateComplexDiffusion_OADA(par=c(5,10), network=simNet1,
transmissionFunction = logisticLearning)
```

#The object thresholdModelSimData then provides the information on each individuals' time of adoption of the behaviour over the social network given the Frequency-dependent contagion rule (see Figure 2D for example).

```r
#Fit a model to the simulated data, assuming the same transmission function
model_threshold<-complexOadaFit(thresholdModelSimData, transmissionFunction = logisticLearning, noParTransFunct=2,startValue = c(1,1),lower=c(0,0))

#Obtain parameter estimates
data.frame(variable=model_threshold@varNames,MLE=model_threshold@outputPar, SE=model_threshold@se)
```

Obtaining:
```
                   variable       MLE         SE
1 1 Transmission parameter 1  5.601823  0.3244505
2 2 Transmission parameter 2 21.593520 17.8873635
```

We can see that *a* has been estimated close to the value used to generate the data, whereas *c* is estimated as higher (see below for an investigation of the performance of this model). Code for obtaining 95% confidence intervals is provided in the Supplementary Material example script "Threshold function example.R". These were found to be 4.8 – 6.2 for the location of the threshold and 4.4 – 126 for the size parameter.

```r
#Get AICc for the model
model_threshold@aicc
#[1] 710.159

#Create a standard nbdaData object
nbdaData1<-nbdaData(label="thresholdStandard",assMatrix=simNet1,orderAcq = thresholdModelSimData@orderAcq)
#Fit the model to the data
model_standard <-oadaFit(nbdaData1)
#Get the AICc
model_standard@aicc
#[1] 726.8548

model_asocial<-oadaFit(nbdaData1,type="asocial")
model_asocial @aicc
#[1] 727.4788
```

**Supplementary Methods B: Description of Supplementary R Scripts**

**All the below scripts are available at: https://github.com/whoppitt/complexNBDA**

Along with the walk-though examples (Methods Section 2; Supplementary Methods 1), we also provide the R Scripts for running each of the social transmission rules (Figure 2B,C,D) in the following scripts:

**(2A)** Script "Proportion rule example.R" – For building, simulating and testing the proportional social transmission rule model (See Main Text Methods Section 1i, 2, & 3i – and Figure 1B & 2B)

**(2B)** Script "Frequency dependent example.R" – For building, simulating and testing the Frequency-dependent social transmission rule model (See Main Text Methods Section 1ii & 3ii – and Supplementary Methods 1A - and Figure 2C). Also includes the script for comparing different social transmission rule models against one another, and for using simulations for adjusting confidence intervals to appropriate level.

**(2C)** Script "Threshold example.R" – For building, simulating and testing the threshold social transmission rule model (See Main Text Methods Section 1iii & 3iii – and Supplementary Methods 1B - and Figure 1C & 2D). Also includes code for using multiple simulations for testing the performance of the threshold model over different datasets, and for interpreting and adjusting CI's to appropriate level.

We also provide the base scripts for the package i.e. the complexNDBA model and the associated functions (object creation, simulations, likelihood)

**(2D)** Script "complexOadaFit.R" – Implementing the complexNBDA model fitting

**(2E)** Script "complexOadaLikelihood.R" – Implementing likelihood calculation for complexNBDA

**(2F)** Script "complexProfLikCI.R" – determining profile of likelihood for parameters (from specified complexOadaFit model)

**(2G)** Script "complexNBDAdata.R" – Creating objects of class complexNBDAdat

**(2H)** Script "simulateComplexDiffusion.R" – Simulating network diffusions from a specified transmission function

**SUPPLEMENTARY TABLES**

**Table S1.** Testing the performance of the frequency-dependent model for estimating 95% confidence intervals of the frequency-dependent parameter *f*. Values show the percentage of cases the true value of the frequency-dependent parameter *f* lays within the 95% confidence intervals for estimated *f* as a function of the true value of the size parameter *s* (rows) and *f* (columns).

| True value of *s* | % of cases within 95% C.I. | | | |
|---|---|---|---|---|
| | f= 1 | f= 2 | f= 3 | f= 5 |
| 5 | 94.5 | 94.5 | 93.7 | 93.9 |
| 10 | 94.7 | 93.1 | 93.5 | 93.1 |
| 30 | 95.7 | 93.2 | 92.6 | 94.3 |

**Table S2.** Testing the performance of the frequency-dependent model for estimating 95% confidence intervals of the size parameter *s*. Values show the percentage of cases the true value of the size parameter *s* lays within the 95% confidence intervals for estimated *s* as a function of the true value of the size *s* (rows) and the Frequency-dependent parameter *f* (columns).

| True value of *s* | % of cases within 95% C.I. | | | |
|---|---|---|---|---|
| 0 | 92.1 | | | |
| | f= 1 | f= 2 | f= 3 | f= 5 |
| 5 | 86.5 | 84.4 | 84.4 | 83.5 |
| 10 | 85.7 | 82.9 | 82.6 | 81.3 |
| 30 | 84.9 | 85.1 | 82.3 | 83.6 |

**Table S3.** Testing the performance of the threshold rule model for estimating 95% confidence intervals of the location parameter *a*. Values show the percentage of cases the true value of the location parameter *a* lays within the 95% confidence intervals for estimated *a* as a function of the true value of the size parameter *c* (rows) and *a* (columns).

| True value of *c* | % of cases within 95% C.I. | | | |
|---|---|---|---|---|
| | a= 2.5 | a= 5 | a= 10 | a= 15 |
| 5 | 91.2 | 93.3 | 93.8 | 93.9 |
| 10 | 89.4 | 92.9 | 94.8 | 95.0 |

**Table S4.** Testing the performance of the threshold rule model for estimating 95% confidence intervals of the size parameter *c*. Values show the percentage of cases the true value of the size parameter *c* lays within the 95% confidence intervals for estimated *c* as a function of the true value of the *c* (rows) and location parameter *a* (columns).

| True value of *c* | % of cases within 95% C.I. | | | |
|---|---|---|---|---|
| 0 | 90.6 | | | |
| | a= 2.5 | a= 5 | a= 10 | a= 15 |
| 5 | 94.8 | 95.4 | 95.1 | 95.2 |
| 10 | 95.2 | 95.2 | 96.9 | 94.5 |

**SUPPLEMENTARY VIDEO 1 DESCRIPTION**

Supplementary Video 1 shows an animated version of the contagions in relation to Figure 2. The video runs from time-step 0 (1 informed node) to time-step 60 (61 informed nodes) over each contagion type (i.e. the same as Figure 2). The panels show each scenario **(A)** Simple contagion **(B)** Proportional rule **(C)** Frequency-dependent rule, and **(D)** Threshold rule. The 't=' mark denotes the time-step within the contagion (i.e. number of individuals adopted). For display purposes, time-steps 0 to 9 are shown at a 1 time-step per frame rate, 10-20 is shown every 2 time steps, 21-39 is shown every 3 time-steps and 40-60 is shown every 4 time steps. Across all panels the exact same network structure is maintained, and each has the same start node. The node size shows the number of social ties each individual has, while node shape shows individuals' state: circles denote uninformed individuals and the green squares represent 'informed' individuals, with the shade of the green showing the time since adopting the behaviour (darker = earlier adoption). The adjoining lines show the social ties between individuals, with links to informed individuals as green lines and links between uninformed individuals as grey lines. The colour of the circles (uninformed nodes) represent the relative likelihood of each uninformed individual's chance of being the next individual to adopt the behaviour given their social links and given the contagion rule at play within each panel. White nodes show individuals with no links to informed individuals at that time step.